\documentclass[letterpaper,12pt]{article}
\usepackage{amsmath,amsfonts}
\usepackage{graphicx}

\providecommand{\manif}[1]{\ensuremath{\mathbb{#1}}}
\providecommand{\dX}{\ensuremath{\dot X}}
\providecommand{\D}{\mathcal{D}}
\providecommand{\tg}{\tilde \Gamma}
\providecommand{\pa}{\partial_\alpha}
\providecommand{\pb}{\partial_\beta}
\providecommand{\pc}{\partial_\gamma}
\providecommand{\pd}{\partial_\delta}
\providecommand{\ford}[1]{\underset{(1)}{#1}}
\providecommand{\sord}[1]{\underset{(2)}{#1}}
\providecommand{\tord}[1]{\underset{(3)}{#1}}
\providecommand{\N}{\mathcal{N}}

\title{The Method of Geodesic Expansions\\
and its Application to the Semiclassical Sum over Immersed Manifolds}
\author{Wolfgang M\"uck\thanks{e-mail: wmueck@sfu.ca}\\
\small Department of Physics, Simon Fraser University,
Burnaby, B.C., V5A 1S6 Canada}

\begin{document}
\maketitle

\begin{abstract}
The method of geodesic expansions is systematically explained. Based on the Haar measures of the group of geodesic expansions the semiclassical sum over immersed manifolds is constructed. Gauge fixing is performed via the Faddeev Popov method.
\end{abstract}

\renewcommand{\baselinestretch}{1.2} \normalsize

\section{Introduction}
Immersed Manifolds play an important role in the description of relativistic objects (point particles, strings, etc.). Namely, an object of dimension $n$ sweeps out in time an $n+1$ dimensional ``world'' manifold $\manif{W}$, which is immersed in the surrounding space-time. The quantization of such objects is formally described by the Euclidean partition function \cite{Polyakov}
\begin{equation}
\label{intro:Z}
  Z = \sum_{\{X\}} \mathrm{e}^{-S} = \sum_{\mathrm{top.}} \int \frac{\D X}{\D f} \mathrm{e}^{-S}.
\end{equation}
All of these theories have in common the fact that the canonically conjugate variables, say $X^\mu$ and $P_\mu$, satisfy constraints, which originate in the reparametrization invariance of $\manif{W}$. One usually adds the constraints to the Hamiltonian via Lagrange multipliers \cite{Hwang} yielding the same classical physics. The quantization of the free relativistic particle \cite{Kleinert} and of the free bosonic string \cite{polyakov1} have elegantly been performed in this way. 
A more sophisticated method of quantization of systems with constraints is the BRST procedure \cite{Henneaux}. 

Despite these successes the question is justified, if a quantization is feasible without introducing the constraints, namely by considering only the physical degrees of freedom. As a first step one could attempt a semiclassical treatment, which would naturally allow for a curved space-time. Attempts in this direction have been made \cite{Vish96}, but suffer from an unsatisfactory treatment of the path integral measure. 

The purpose of this article is to lay a foundation for the semiclassical treatment of systems described by immersions $\manif{W}$. It is organized as follows. Sec.~\ref{geod} deals with the method of geodesic expansions. We shall establish the group structure of the geodesic expansions in Sec.~\ref{geod:princ}, construct the Haar measures corresponding to this group in Sec.~\ref{geod:integ} and find the significance of the left invariant Haar measure in Sec.~\ref{geod:diff}. The geodesic expansion has long been used in the literature (e.g.\ Eqn.\ \eqref{geod:rmeas} is used in \cite{Gavazzi}), but is usually not studied on its own  because of its conceptual simplicity. Thus, Sec.~\ref{geod} can be regarded as a systematic review of the method. In Sec.~\ref{appl} we will apply it to the semiclassical treatment of the sum over $\manif{W}$. Of particular importance is Sec.~\ref{appl:diff}, which describes in detail, how diffeomorphisms of $\manif{W}$ are represented in the method of geodesic expansions and which thereby lays the foundation for the explicit gauge fixing by means of the Faddeev Popov method in Sec.~\ref{appl:gauge}. Finally, Sec.~\ref{concl} contains conclusions. 

\section{The Method of Geodesic Expansions}
\label{geod}
\subsection{Principal Idea and Group Structure}
\label{geod:princ}
Consider the points $X$ of a manifold \manif{M} with Riemannian structure, i.e.\ $X$ has components $X^a$, where $a$ belongs to an arbitrary set of indices (discrete as in usual Riemannian spaces or continuous as in functional spaces) \cite{Basler}. Furthermore, a metric tensor $h_{ab}$ exists such that 
\begin{equation}
\label{geod:metric}
  ||\delta X||^2 = h_{ab}[X] \delta X^a \delta X^b.
\end{equation}
The metric connections (Christoffel symbols) and curvature tensor are defined as usual \cite{MTW}
\begin{align}
\label{geod:conn}
  \Gamma^a_{bc} &= \frac{1}{2} h^{ad} \left(h_{db,c}+h_{dc,b}-h_{bc,d}\right), \\
\label{geod:curv}
  R^a_{\;bcd} &= \Gamma^a_{bd,c} - \Gamma^a_{bc,d} + \Gamma^a_{ce}\Gamma^e_{bd} - \Gamma^a_{de}\Gamma^e_{bc}.
\end{align}

Now consider two points $X_0$ and $X_1$ in \manif{M}, which are (in some intuitive sense) close to each other and which therefore can be connected by a unique geodesic. Let $\theta$ be the affine parameter of this geodesic, i.e.\ the geodesic equation is 
\begin{equation}
\label{geod:geod}
  \frac{\partial^2}{\partial \theta^2} X^a(\theta) + 
  \Gamma^a_{bc} \frac{\partial}{\partial \theta} X^b \frac{\partial}{\partial \theta} X^c = 0,
\end{equation}
and let $\theta$ be scaled such that 
\[ X^a(0) = X^a_0 \quad \text{and} \quad X^a(1) = X^a_1. \]
Thus, the parameter of the geodesic connecting $X_0$ and $X_1$ is uniquely specified.

The aim is to expand functionals $F[X]$ in a power series in the parameter $\theta$ in order to obtain an approximation for $F[X_1]$. Let us start with $X_1$ itself. The Taylor series naturally begins as
\begin{align}
\label{geod:exp1}
  X^a_1 &=X^a_0 + \left. \left\{ \frac{\partial}{\partial \theta} X^a(\theta) +
  \frac{1}{2} \frac{\partial^2}{\partial \theta^2} X^a (\theta) +
  \frac{1}{6} \frac{\partial^3}{\partial \theta^3} X^a (\theta) \right\} \right|_{\theta=0} +\cdots.\\
\intertext{However, using the geodesic equation \eqref{geod:geod} this can be rewritten as} 
\label{geod:exp2}
  X^a_1 &\simeq X^a_0 + \dX^a - \frac{1}{2} \Gamma^a_{bc} \dX^b \dX^c + 
  \frac{1}{6} (-\Gamma^a_{bc,d} + 2\Gamma^a_{de}\Gamma^e_{bc}) \dX^b \dX^c \dX^d
\end{align}
In \eqref{geod:exp2} all quantities on the r.h.s.\ are evaluated at $\theta=0$ and 
\[ \dot X = \left.\frac{\partial}{\partial \theta} X\right|_{\theta=0}.\]
For the sake of brevity, the conventional dots for higher order terms are henceforth omitted, the symbol $\simeq$ being used instead to indicate the approximation.

Equation \eqref{geod:exp2} shows the great virtue of the geodesic expansion, namely that all higher order derivatives of $X(\theta)$ can be expressed in terms of $\dX$ by means of the geodesic equation. The vector $\dX$, being equal to $\delta X$ in a flat space ($X_1 = X_0 +\delta X$), can quite generally be regarded as the expansion parameter. We will henceforth say that the geodesic expansion maps $X_0$ into $X_1$ and denote this mapping by $\dX: X_0 \rightarrow X_1$.   

There is a different interpretation of the quantities $\dX^a$. Indeed, eqn.\ \eqref{geod:exp2} shows that they are Riemannian normal coordinates of a neighbourhood of $X_0$ \cite{Petrov}. However, we will not consider them as such, but continue to treat them as the components of the vector $\dX$. A Riemannian normal coordinate system will nevertheless be useful for simplifying our calculations. 

Let us establish now that the geodesic expansions introduced above form a group. For this purpose, $\dX$ has to be considered as a vector field $\dX[X]$ with $\dX[X_0]$ being the special value, which we met so far. What about the group axioms? Clearly an identity exists, given by $\dX=0$, and the inverse of any transformation can be found by inverting \eqref{geod:exp2}. We will now show that the product property of groups is also satisfied. This calculation is best performed order by order.

Consider two expansions, $\dX_1$ and $\dX_2$ such that $\dX_1: X_0 \rightarrow X_1$ and $\dX_2: X_1 \rightarrow X_2$. Assume now that the product property is satisfied, i.e.\ there exists an expansion $\dX$ such that $\dX=\dX_2 \circ \dX_1: X_0 \rightarrow X_2$. Consider the final point $X_2$, which is to first order explicitely given by 
\begin{equation}
\label{geod:prod1}
  \ford{X_2^a} = \ford{X_1^a} + \dX_2^a[X_1] 
  = X_0^a + \dX_1^a + \dX_2^a, 
\end{equation}
where $\dX_{1,2}$ denotes $\dX_{1,2}[X_0]$ for brevity and the underset numbers represent the order in $\dX$, up to which the quantity is to be evaluated. Comparing \eqref{geod:prod1} with \eqref{geod:exp2} we can read off
\begin{equation}
\label{geod:fprod}
  \ford{\dX^a} = \dX_1^a + \dX_2^a.
\end{equation}
Continuing with the second order, we find
\begin{align}
\notag 
  \sord{X_2^a} &= \sord{X_1^a} + \dX_2^a[X_1] - \frac{1}{2} \Gamma^a_{bc}[X_1] \dX_2^b[X_1] \dX_2^c[X_1] \\
\notag
  &= X_0^a + \ford{\dX^a} + \dX_1^b \dX_{2\,;b}^a - \frac{1}{2} \Gamma^a_{bc} \ford{\dX^b} \ford{\dX^c}. \\
\intertext{Again, comparing this with \eqref{geod:exp2} we obtain}
\label{geod:sprod}
  \sord{\dX^a} &= \dX_1^a + \dX_2^a + \dX_1^b \dX_{2\,;b}^a.
\end{align}
The calculation of the third order proceeds in the same fashion, but is naturally more involved. Hence we will give only the result,
\begin{equation}
\label{geod:tprod}
  \tord{\dX^a} = \dX_1^a + \dX_2^a + \dX_1^b \dX_{2\,;b}^a + \frac{1}{2} \dX_1^b \dX_1^c \dX_{2\,;bc}^a +\frac{1}{3} R^a_{\;bcd} \left( \dX_2^b + \frac{1}{2} \dX_1^b \right) \dX_2^c \dX_1^d.
\end{equation}
It is straightforward to check from \eqref{geod:tprod} that the associative law is also satisfied. Therefore the geodesic expansions form a group. 

\subsection{Integration Measures}
\label{geod:integ}
The next task is to develop an approximate integration scheme based on the geodesic expansion. It will turn out that the covariant integration measure in $\manif{M}$ naturally leads to the right-invariant Haar measure \cite{Cornwell1} of the group of geodesic expansions. The left-invariant Haar measure will be constructed, but its interpretation postponed until section \ref{geod:diff}. 

Consider the covariant integration in \manif{M}. The integration measure $\D X$ is explicitely given by \cite{Basler}
\begin{equation}
\label{geod:meas}
  \D X = \prod_a [dX^a] \sqrt{|\det h_{ab}[X]|}.
\end{equation}

In a semiclassical approximation one can regard the integration variable $X$ as a variation of a (classical) background $X_0$ and use the geodesic expansion to change integration variables from $X$ to $\dX$. This can be done conventionally as attempted in \cite{Bardakci2}, but is easiest performed using Riemannian normal coordinates \cite{Petrov} $Y^a$ of the neighbourhood of the origin $X_0$. On one hand, the integration measure is independent of the coordinate system, i.e.\
\begin{equation}
\label{geod:measinv}
  \D X = \prod_a [dX^a] \sqrt{|\det h^X_{ab}[X]|} = \D Y = \prod_a [dY^a] \sqrt{|\det h^Y_{ab}[Y]|},
\end{equation}
where $h^X$ and $h^Y$ denote the metrics with respect to the appropriate coordinates. On the other hand, the metric tensor $h_{ab}^Y$ is given by a power series in the normal coordinates $Y$ with all coefficcients being tensors evaluated at the origin \cite{Petrov},
\begin{equation}
\label{geod:hexp}
  h^Y_{ab}[Y] = h^Y_{ab}[0] - \frac13 R^Y_{acbd} Y^c Y^d - \frac16 R^Y_{acbd;e} Y^c Y^d Y^e + \cdots.
\end{equation}
Note that despite the coefficients being tensors \eqref{geod:hexp} is not a tensor equation. Using $\det A = \exp \mathrm{Tr} \ln A$, we obtain from \eqref{geod:hexp} 
\begin{equation}
\label{geod:sqrthexp}
  \sqrt{|h^Y[Y]|} \simeq \sqrt{|h^Y[0]|} \exp \left( -\frac16 R_{ab} Y^a Y^b \right). 
\end{equation}
All we have to do now is realize that, as mentioned above, the coordinates $Y^a$ are identical with the vector components $\dX^a$ and that tensor components with respect to normal coordiantes $Y$ coincide with those with respect to $X$ at the origin. Hence, we find the expression for the integral measure
\begin{equation}
\label{geod:rmeas}
  \D X = \D_R \dX \simeq \sqrt{|h[X_0]|} \prod_a[d\dX^a] 
  \exp \left(-\frac{1}{6} R_{ab}\dX^a \dX^b \right),
\end{equation}
which is covariant, i.e.\ holds in any coordinate system.

The same result would, of course, be obtained by a conventional change of variables. It should be compared to the expression derived in \cite{Bardakci2}, which is non-covariant. The mistake in \cite{Bardakci2} is that only the Jacobian for the change of differentials is considered, but the $\sqrt{h}$ term is not included, which in fact exactly cancels the non-covariant terms of the Jacobian.

By construction, $\D_R \dX$ is the approximate measure of integration over the manifold $\manif{M}$. Let us establish now that it is also the right-invariant Haar measure of the group of geodesic expansions, this being the reason for denoting it with the subscript $R$. 

In order to show the right-invariance of the integral measure \eqref{geod:rmeas}, we consider the measure $\D_R \dX_2$ and change variables to $\dX = \dX_2 \circ \dX_1$. We obtain from \eqref{geod:tprod} to second order
\begin{gather}
\notag
  \frac{\delta \dX^a}{\delta \dX_2^b} \simeq \delta^a_b \left( 1 + \dX_1^c \nabla_c + \frac{1}{2} \dX_1^c \dX_1^d \nabla_c \nabla_d \right) 
  + \frac{1}{3} R^a_{\;cbd} \left( \dX_2^c + \frac{1}{2} \dX_1^c \right) \dX_1^d + \frac{1}{3} R^a_{\;bcd} \dX_2^c \dX_1^d. \\
\intertext{The Jacobian of this change of variables is obtained using the fact that the derivative operator is antisymmetric, i.e.\ its functional trace vanishes. One finds}
\label{geod:jac3}
  \det \left( \frac{\delta \dX^a}{\delta \dX_2^b} \right) 
  \simeq \exp\left( \frac{1}{3} R_{ab} \dX_2^a \dX_1^b + \frac{1}{6} R_{ab} \dX_1^a \dX_1^b \right). \\
\intertext{Hence}
\label{geod:jac4}
  \det \left( \frac{\delta \dX_2^a}{\delta \dX^b} \right)  
  \exp \left(-\frac{1}{6} R_{ab}\dX_2^a \dX_2^b \right) \simeq
  \exp \left(-\frac{1}{6} R_{ab}\dX^a \dX^b \right)
\end{gather}
and from \eqref{geod:rmeas} follows that
\begin{equation}
\label{geod:rinv}
  \D_R \dX = \D_R \left( \dX_2 \circ \dX_1 \right) = \D_R \dX_2.
\end{equation}
This establishes the result that the integral measure \eqref{geod:rmeas} is the right-invariant Haar measure of the group of geodesic expansions.

The left-invariant Haar measure differs, as in the general case \cite{Cornwell1}, from the right-invariant Haar measure. We shall now construct it. Starting again from \eqref{geod:tprod}, we find
\begin{align}
\notag
  \frac{\delta \dX^a}{\delta \dX_1^b} &\simeq \delta^a_b + \dX_{2;b}^a + \frac12 \dX_1^c \left( \dX_{2;bc}^a + \dX_{2;cb}^a \right) 
  + \frac13 R^a_{\;dcb} \left( \dX_2^d + \frac12 \dX_1^d \right) \dX_2^c + \frac16 R^a_{\;bcd} \dX_2^c \dX_1^d,\\
\intertext{which leads to}
\label{geod:ldet}
  \det \left( \frac{\delta \dX^a}{\delta \dX_1^b} \right) &\simeq \exp \left( \dX^a_{;a} - \dX_{1;a}^a -\frac12 \dX^a_{;b} \dX^b_{;a} + \frac12 \dX_{1;b}^a \dX_{1;a}^b - \frac13 R_{ab} \dX^a \dX^b + \frac13 R_{ab} \dX_1^a \dX_1^b \right),
\end{align}
where $\dX_2$ has been eliminated using \eqref{geod:sprod}. Rewriting \eqref{geod:ldet} as 
\begin{multline}
\notag
  \det \left( \frac{\delta \dX_1^b}{\delta \dX^a} \right) \exp \left( - \dX_{1;a}^a + \frac12 \dX_{1;b}^a \dX_{1;a}^b + \frac13 R_{ab} \dX_1^a \dX_1^b \right)\\
  \simeq \exp \left( - \dX^a_{;a} + \frac12 \dX^a_{;b} \dX^b_{;a} + \frac13 R_{ab} \dX^a \dX^b \right)
\end{multline}
it is realized that the integral measure 
\begin{equation}
\label{geod:lmeas}
 \D_L \dX \simeq \sqrt{|h[X_0]|} \prod_a [d\dX^a] \exp \left( - \dX^a_{;a} + \frac12 \dX^a_{;b} \dX^b_{;a} + \frac13 R_{ab} \dX^a \dX^b \right)
\end{equation}
is the left-invariant Haar measure, i.e.\ it satisfies
\begin{equation}
\label{geod:linv}
  \D_L \dX = \D_L \left( \dX_2 \circ \dX_1 \right) = \D \dX_1.
\end{equation}

In contrast to the right-invariant Haar measure of the group of geodesic expansions, $\D_R \dX$, which was obtained naturally from the measure of covariant integration over the manifold $\manif{M}$, the integral measure $\D_L \dX$ has been constructed from the knowledge of the product of two group elements. However, it will turn out in the next section that it is intimately linked to the integration measure over the group of reparametrizations of the manifold $\manif{M}$.

\subsection{Integral over Diffeomorphisms}
\label{geod:diff}
Let $\manif{M}$ now be a finite dimensional manifold of dimension $n$. Consider two different coordinate charts on $\manif{M}$, e.g.\ with coordinates $X$ and $Y$. The metric tensors with respect to these coordinates are related by 
\begin{equation}
\label{geod:hy}
  h^Y_{ab} = h^X_{cd}\; \frac{\partial X^c}{\partial Y^a} \frac{\partial X^d}{\partial Y^b}.
\end{equation}

The functions $Y(X)$ reparametrize the manifold $\manif{M}$ and thus represent a diffeomorphism on $\manif{M}$. As is well known, the diffeomorphisms form a group. Polyakov \cite{Polyakov} provides the unique metric in the functional space of diffeomorphisms, which is invariant under both, left and right group multiplications. It is given by
\begin{align}
\label{geod:diffmet1}
  ||\delta Y||^2 &= \int d^n X \sqrt{h^X[X]} \; h^X_{ab}[X] \delta Y^a \left(Y^{-1}(X)\right) \delta Y^b \left(Y^{-1}(X)\right).\\
\intertext{Relabelling $X\rightarrow Y(X)$ and using the invariance of the volume element, \eqref{geod:diffmet1} becomes}
\label{geod:diffmet2}
  ||\delta Y||^2 &= \int d^n X \sqrt{h^X[X]} \; h^Y_{ab}[Y(X)] \delta Y^a(X) \delta Y^b(X).
\end{align}
The integration measure in the space of diffeomorphisms is readily obtained from \eqref{geod:diffmet2} as 
\begin{equation}
\label{geod:measy}
  \D Y = \prod_X \left\{ \left(h^X[X]\right)^\frac{n}{4} d^n Y(X) \sqrt{h^Y[Y(X)]}\right\}.
\end{equation}

We are now in a position to make the connection to the method of geodesic expansions. Namely, consider the diffeomorphism $Y(X)$ as generated by the formula (cf.\ Eqn.\ \eqref{geod:exp2})
\begin{equation}
\label{geod:expy}
  Y^a(X) \simeq X^a + \dX^a - \frac{1}{2} \Gamma^a_{bc} \dX^b \dX^c + 
  \frac{1}{6} (-\Gamma^a_{bc,d} + 2\Gamma^a_{de}\Gamma^e_{bc}) \dX^b \dX^c \dX^d,
\end{equation}
where the argument $X$ has been omitted on the r.h.s. Changing integration variables to $\dX$ we find the Jacobian (note the non-covariant terms)
\begin{equation}
\label{geod:jac5}
  \det \left(\frac{\delta Y^a}{\delta \dX^b} \right) \simeq
  \exp \left( -\frac{1}{6} R_{ab} \dX^a \dX^b - \Gamma^a_{ab} \dX^b - \frac{1}{2} \Gamma^c_{cb;a} \dX^a \dX^b \right).
\end{equation}
The $\sqrt{h}$ term will again cancel the non-covariant terms, but the final result will differ from \eqref{geod:rmeas}, because we are dealing with a passive transformation. Thus, instead of expanding $\sqrt{h}$ we have to use \eqref{geod:hy}, which yields
\begin{equation}
\label{geod:hyexp}
  \sqrt{|h^Y[Y(X)]|} = \sqrt{|h^X[X]|} \left| \det \left( \frac{\delta Y^a}{\delta X^b} \right) \right|^{-1}.
\end{equation}
From \eqref{geod:expy} we find
\begin{equation}
\notag
  \frac{\delta Y^a}{\delta X^b} \simeq \delta^a_b +\dX^a_{,b} -\frac12 \Gamma^a_{cd,b} \dX^c \dX^d - \Gamma^a_{cd} \dX^c_{,b} \dX^d,
\end{equation}
which gives
\begin{equation}
\label{geod:hydet}
  \det \left( \frac{\delta Y^a}{\delta X^b} \right) \simeq \exp \left( -\Gamma^a_{ab} \dX^b -\frac12 \Gamma^a_{ab;c} \dX^b \dX^c + \dX^a_{;a} - \frac12 \dX^a_{;b} \dX^b_{;a} - \frac12 R_{ab} \dX^a \dX^b \right).
\end{equation}
Hence, the final result for the approximate measure of integration over the diffeomorphisms of $\manif{M}$ is the covariant expression
\begin{equation}
\label{geod:measy2}
  \D Y \simeq \prod_X \left[ \left(h^X[X] \right)^\frac{n+2}{4} d^n \dX(X) \exp \left( - \dX^a_{;a} + \frac12 \dX^a_{;b} \dX^b_{;a} + \frac13 R_{ab} \dX^a \dX^b \right) \right].
\end{equation}
Comparing \eqref{geod:measy2} to \eqref{geod:lmeas} we see that 
\begin{equation}
\label{geod:measyl}
  \D Y = \prod_X \left(h^X[X]\right)^\frac{n}{4} \D_L \dX[X],
\end{equation}
i.e.\ the left invariant Haar measure of the group of geodesic expansions is closely related to the measure of integration over the diffeomorphisms.

\section{The Semiclassical Sum over Immersed Manifolds}
\label{appl}
The method of geodesic expansions shall now be applied to the semiclassical treatment of the sum over immersed manifolds. Specifically, an explicit expression for the integral on the r.h.s.\ of \eqref{intro:Z} (with the topology given by the topology of the chosen classical configuration) will be derived. In order to do so, we first need to review the geometrical relations satisfied by immersed manifolds. This will be done in Sec.~\ref{appl:geom}. After showing in Sec.~\ref{appl:riem} that the space of immersions has a Riemannian structure, the method of geodesic expansions becomes directly applicable to the problem.

\subsection{Geometrical Relations for Immersed Manifolds}
\label{appl:geom}
Let $X$ be a smooth mapping of a $d$-dimensional parameter manifold \manif{P} into the Riemannian space-time manifold $\manif{S}$, 
\[ X: \sigma \rightarrow X(\sigma); \quad \manif{P} \rightarrow \manif{W} \subset \manif{S}. \]
Clearly, the manifolds $\manif{P}$ and $\manif{W}$ are isomorphic and $\manif{W}$, which represents the ``world line'', ``world sheet'', etc. for $d=1,2,\ldots$, possesses the induced metric
\begin{equation}
  \label{geom:gind}
    g_{\alpha\beta} = \pa X^\mu \pb X^\nu h_{\mu\nu},
  \end{equation}
where $h_{\mu\nu}$ is the metric on $\manif{S}$ ($\alpha,\beta=1,2,\ldots,d$; $\mu,\nu=1,2,\ldots,D$). The vectors $\pa X^\mu$ provide a local coordinate frame on $\manif{W}$ and form its tangent vector bundle. In addition to these we introduce a system of $D-d$ normal vectors $N_i^\mu$ ($i=d+1,\ldots,D$), which satisfy, together with the tangent vectors, the orthogonality and completeness relations
\begin{align}
\label{geom:ntort}
  N_i^\mu \pa X^\nu h_{\mu\nu} &= 0, \\
\label{geom:nnort}
  N_i^\mu N_j^\nu h_{\mu\nu} &= \epsilon(i) \delta_{ij}, \\
\label{geom:compl}
  g^{\alpha\beta} \pa X^\mu \pb X^\nu + 
  \epsilon(i) \delta^{ij} N_i^\mu N_j^\nu  &= h^{\mu\nu}.
\end{align}
Here $\epsilon(i) \delta_{ij}$ is the pseudo Euclidean metric in the normal bundle with $\epsilon(i)=\pm 1$ according to the signatures of the metrics on $\manif{W}$ and $\manif{S}$. In the following we will assume that $\epsilon(i)=1\;\forall i$.

Because $\manif{W}$ is immersed in $\manif{S}$, its geometrical properties are determined by structure equations \cite{Eisenhart}. First, there is the equation of Gauss 
\begin{equation}
\label{geom:gauss1}
  \nabla_\alpha \pb X^\mu \equiv 
  \pa \pb X^\mu + 
  \tg^\mu_{\nu\rho} \pa X^\nu \pb X^\rho
  - \Gamma^\gamma_{\alpha\beta} \pc X^\mu 
  = H^i_{\alpha\beta} N_i^\mu \equiv H^\mu_{\alpha\beta},
\end{equation}
which defines the second fundamental forms $H^i_{\alpha\beta}$. The quantity $H^\mu = \frac12 H^{\mu\alpha}_\alpha$ is the mean curvature vector. In Eqn.\ \eqref{geom:gauss1} and all subsequent formulae symbols with a tilde represent quantities on $\manif{S}$, those unadorned on $\manif{W}$. Furthermore, the symbol $\nabla$ denotes the covariant derivative with respect to all indices. 

Second, the equation of Weingarten
\begin{equation} 
\label{geom:wein}
  \nabla_\alpha N^{i\mu} \equiv 
  \pa N^{i\mu} + \tg^\mu_{\nu\rho} \pa X^\nu N^{i\rho}
  + A^i_{\,j\alpha} N^{j\mu} 
  = - H^{i\beta}_{\alpha} \pb X^\mu
\end{equation}
introduces the gauge connections in the normal bundle, $A^i_{\,j\alpha}$. 
It can be noted that the gauge connections $A^i_{\,j\alpha}$ are not tensors, because they transform like gauge fields under local $SO(D-d)$ rotations in the normal bundle. However, the associated field strength (also called the normal curvature tensor)
\begin{equation}
\label{geom:ffield}
  F^i_{\;j\alpha\beta} = 
  \pa A^i_{\,j\beta} - \pb A^i_{\,j\alpha}
  + A^i_{\,k\alpha} A^k_{\,j\beta} - A^i_{\,k\beta} A^k_{\,j\alpha}
\end{equation}
is a genuine tensor with respect to all indices.

Further imortant equations can be derived by taking covariant derivatives of \eqref{geom:gauss1} and \eqref{geom:wein}. They are, respectively, the equations of Gauss, Codazzi, and Ricci \cite{Eisenhart}
\begin{align}
\label{geom:gauss2}
  \tilde R_{\mu\nu\lambda\rho} \pa X^\mu \pb X^\nu
  \pc X^\lambda \pd X^\rho &= R_{\alpha\beta\gamma\delta} 
  + H^i_{\alpha\delta} H_{i\beta\gamma}
  - H^i_{\alpha\gamma} H_{i\beta\delta} \\
\label{geom:codazzi}
  \tilde R_{\mu\nu\lambda\rho}
  \pa X^\mu \pb X^\nu \pc X^\rho N^{i\lambda} &= \nabla_\alpha H^i_{\beta\gamma} - \nabla_\beta H^i_{\alpha\gamma} \\
\label{geom:ricci}
  \tilde R_{\mu\nu\lambda\rho} \pa X^\mu \pb X^\nu 
  N^{i\lambda} N_j^\rho &= F^i_{\;j\alpha\beta}
  - H^{i\gamma}_{\alpha} H_{j\gamma\beta} 
  + H^{i\gamma}_{\beta} H_{j\gamma\alpha}.
\end{align}

\subsection{The Riemannian Structure of the Space of Immersions}
\label{appl:riem}
Let us start by identifying the Riemannian structure of the space $\manif{M}$ of mappings $X: \manif{P} \rightarrow \manif{W}$. The coordinates $X^a$ of $\manif{M}$  are labelled by both, the space-time index $\mu$ and the continuous index $\sigma$. We shall write $a=(\mu\sigma)$. The norm on $\manif{M}$ is given by \cite{Basler}
\begin{equation}
\label{appl:norm}
  ds^2 = \frac{1}{\N} \int d^d\sigma \sqrt{|g(\sigma)|}\; h_{\mu\nu}[X(\sigma)] dX^\mu(\sigma) dX^\nu(\sigma),
\end{equation}
where $g$ is the determinant of the metric of $\manif{P}$ and $\N$ is a constant of dimension (length)$^d$. From \eqref{appl:norm} the metric tensor $h_{ab}$ of \manif{M} (cf.\ Sec.\ \ref{geod}) can be identified as 
\begin{equation}
\label{appl:metr}
  h_{(\mu\sigma)(\nu\sigma')} = \N^{-1} h_{\mu\nu}[X(\sigma)] \sqrt{|g(\sigma)|}\; \delta(\sigma,\sigma')
\end{equation}
with $\delta(\sigma,\sigma')$ being the Dirac delta functional on \manif{P} \footnote{We write $\delta(\sigma,\sigma')$ instead of the more familiar $\delta(\sigma-\sigma')$, because in general there is no translational symmetry on $\manif{P}$.}. We also find
\begin{align}
\label{appl:metr2}
  h^{(\mu\sigma)(\nu\sigma')} &= \N h^{\mu\nu}[X(\sigma)] |g(\sigma)|^{-\frac12} \delta(\sigma,\sigma'), \\
\label{appl:metr3}
  \delta^{(\mu\sigma)}_{(\nu\sigma')} &= \delta^\mu_\nu \delta(\sigma,\sigma'), \\
\label{appl:metr4}
  \delta^{(\mu\sigma)}_{(\mu\sigma)} &= D \N^{-1} \int d^d \sigma \sqrt{|g(\sigma)|} = D \N^{-1} \mathcal{V}, 
\end{align}
where $\mathcal{V}$ is the volume of $\manif{P}$. The justification of \eqref{appl:metr4} involves specifying the value of $\delta(\sigma,\sigma)$, which is usually assumed to be infinite. However, no contradiction arises from making it finite, since this single value does not change the integral property of the delta distribution
\begin{equation}
\label{appl:deltaprop}
  \int d^d \sigma' \, \delta(\sigma,\sigma') f(\sigma') = f(\sigma).
\end{equation}
In order to identify $\delta(\sigma,\sigma)$, it has to be noted that the delta functional transforms as $\sqrt{|g(\sigma)|}$ under diffeomorphisms of $\manif{P}$ \cite{Basler}. Hence a finite value must have the form
$\delta(\sigma,\sigma) = \mathrm{const.} \sqrt{|g(\sigma)|}$. Identifying the constant with $\N^{-1}$ for the right unit we arrive at \eqref{appl:metr4}.

A differential structure of $\manif{M}$ is provided by the ordinary functional derivative, $Y_{,(\mu\sigma)} = \frac{\delta Y}{\delta X^\mu(\sigma)}$. Using the definitions \eqref{geod:conn} and \eqref{geod:curv} one obtains the metric connections, curvature tensor and Ricci tensor of $\manif{M}$ as
\begin{align}
\label{appl:conn}
  \Gamma^{(\mu\sigma)}_{(\nu\sigma')(\lambda\sigma'')} &= 
  \tg^\mu_{\nu\lambda}[X(\sigma)] \delta(\sigma,\sigma') \delta(\sigma,\sigma''), \\
\label{appl:curv}
  R^{(\mu\sigma)}_{\quad(\nu\sigma')(\lambda\sigma'')(\rho\sigma''')} &=
  \tilde R^\mu_{\;\nu\lambda\rho}[X(\sigma)] \delta(\sigma,\sigma') \delta(\sigma,\sigma'') \delta(\sigma,\sigma'''), \\
\label{appl:ricci}
  R_{(\mu\sigma)(\nu\sigma')} &= \N^{-1}
  \tilde R_{\mu\nu}[X(\sigma)] \sqrt{|g(\sigma)|} \delta(\sigma,\sigma').
\end{align}

\subsection{Diffeomorphisms}
\label{appl:diff}
We shall now draw our attention to diffeomorphisms of the immersed manifold. In order to make contact with the method of geodesic expansion, we will consider the field $\dX(\sigma)$ on $\manif{P}$, which forms a map
\[ \dX: \manif{W}_0 \rightarrow \manif{W}, \]
generating the immersed manifold $\manif{W}$ from a (classical or background) manifold $\manif{W}_0$. Semiclassically, this is equivalent to considering the map $X$.

A diffeomorphism will be represented by a transformation $f_\eta: \dX \rightarrow \dX' = f_\eta \circ \dX$, which leaves $\manif{W}$ invariant. Let $f$ be a diffeomorphism of $\manif{P}$. Within the method of geodesic expansion, $f$ should have the form $f(\sigma) = \sigma + \delta\sigma$, where (cf.\ Eqn.\ \eqref{geod:exp2})
\begin{equation}
\label{appl:dels}
  \delta\sigma^\alpha \simeq \eta^\alpha 
  - \frac{1}{2} \Gamma^\alpha_{\beta\gamma} \eta^\beta \eta^\gamma 
  + \frac{1}{6} \left(-\Gamma^\alpha_{\beta\gamma,\delta} + 2 \Gamma^\alpha_{\delta\varepsilon} \Gamma^\varepsilon_{\beta\gamma} \right)
  \eta^\beta \eta^\gamma \eta^\delta.
\end{equation}
The vector field $\eta(\sigma)$ is the generator of this geodesic expansion on $\manif{P}$. Moreover, since $\manif{P}$ and $\manif{W}_0$ are isomorphic, it also generates a geodesic expansion on $\manif{W}_0$ by the prescription $\eta \circ X_0(\sigma) = X_0(f(\sigma))$. 

\begin{figure}
 \begin{center}
  \fbox{\includegraphics[bb=95 172 654 514,clip,width=0.5\textwidth]{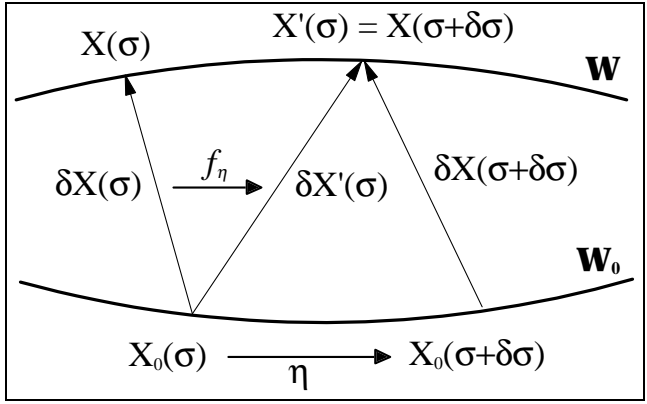}}
 \end{center}
 \caption{Illustration of an automorphism in flat space-time ($\dX=\delta X$). \label{appl:autmorph}}
\end{figure}

Consider now the map $\dX \circ \eta: X_0(\sigma) \rightarrow X(f(\sigma))$. Obviously, its image is $\manif{W}$, i.e.\ identical to the image of $\dX$. Hence, there must exist a map $\dX'$ such that $\dX' = \dX \circ \eta$. In other words, $\eta$ also generates the transformation 
\begin{equation}
\label{appl:f}
  f_\eta: \dX \rightarrow \dX' = f_\eta \circ \dX = \dX \circ \eta. 
\end{equation}
The actions of all these transformations are illustrated in Fig.~\ref{appl:autmorph}.

Let us proceed by calculating the map $\dX'$ in terms of $\dX$ and $\eta$, a task, which is done best order by order similar to the calculation of the product of two geodesic expansions in Sec.~\ref{geod}. To first order we have
\begin{align}
\notag
  \ford{X'}^\mu(\sigma) = \ford{X^\mu}(\sigma+\delta\sigma) &= \ford{X_0^\mu}(\sigma+\delta\sigma)+\ford{\dX^\mu}(\sigma+\delta\sigma)\\
\label{appl:aut1}
  &= X_0^\mu + \eta^\alpha \pa X_0^\mu + \dX^\mu, \\
\intertext{where the argument $\sigma$ has been omitted on the second line. Comparing \eqref{appl:aut1} with}
\notag
  \ford{X'}^\mu &= X_0^\mu + \ford{\dX'}{}^\mu,\\
\intertext{we arrive at}
\label{appl:aut2}
  \ford{\dX'}{}^\mu &= \dX^\mu + \eta^\alpha \pa X_0^\mu.
\end{align}

The second order step starts with
\begin{align}
\notag
  \sord{X^\mu}(\sigma+\delta\sigma) &= \sord{X_0^\mu}(\sigma+\delta\sigma) + \sord{\dX^\mu}(\sigma+\delta\sigma) - \frac{1}{2} \tg^\mu_{\nu\rho} \dX^\nu \dX^\rho \\
\notag
  &= X_0^\mu + \left(\eta^\alpha - \frac{1}{2} \eta^\beta \eta^\gamma \Gamma^\alpha_{\beta\gamma} \right) \pa X_0^\mu + \frac{1}{2} \eta^\alpha \eta^\beta \pa \pb X_0^\mu \\
\notag 
  &\quad + \dX^\mu + \eta^\alpha \pa \dX^\mu - \frac{1}{2} \tg^\mu_{\nu\rho}  \dX^\nu \dX^\rho \\
\label{appl:aut3}
  &= X_0^\mu + \ford{\dX'}{}^\mu + \frac{1}{2} \eta^\alpha \eta^\beta H^i_{\alpha\beta} N_i^\mu + \eta^\alpha \nabla_\alpha \dX^\mu - \frac{1}{2} \tg^\mu_{\nu\rho} \ford{\dX'}{}^\nu \ford{\dX'}{}^\rho. \\
\intertext{As above, comparison of \eqref{appl:aut3} with}
\notag
  \sord{X'}^\mu(\sigma) &= X_0^\mu + \sord{\dX'}{}^\mu - \frac{1}{2} \tg^\mu_{\nu\rho} \ford{\dX'}{}^\nu \ford{\dX'}{}^\rho \\
\intertext{yields}
\label{appl:aut4}
  \sord{\dX'}{}^\mu &= \dX^\mu + \eta^\alpha \pa X_0^\mu + \eta^\alpha \nabla_\alpha \dX^\mu + \frac{1}{2} \eta^\alpha \eta^\beta H^i_{\alpha\beta} N_i^\mu.
\end{align}

As the third order step follows the same pattern, but is rather lengthy, only the result will be given,
\begin{multline}
\label{appl:aut5}
  \tord{\dX'}{}^\mu = \dX^\mu + \eta^\alpha \pa X_0^\mu + \eta^\alpha \nabla_\alpha \dX^\mu + \frac{1}{2} \eta^\alpha \eta^\beta H^i_{\alpha\beta} N_i^\mu + \frac{1}{2} \eta^\alpha \eta^\beta \nabla_\alpha \nabla_\beta \dX^\mu \\
  + \frac{1}{6} \eta^\alpha \eta^\beta \eta^\gamma \nabla_\alpha \left( H^i_{\beta\gamma} N_i^\mu \right) + \frac{1}{3} \tilde R^\mu_{\;\nu\lambda\rho} \left( \dX^\nu + \frac{1}{2} \eta^\alpha \pa X_0^\nu \right) \dX^\lambda \eta^\beta \pb X_0^\rho.
\end{multline}

Let us conclude this subsection by specifying the semiclassical integral measure for the integration over the manifold $\manif{M}$ of immersions $X$. It should be an integral over the field $\dX$ and invariant under $f_\eta$. However, because we can write $f_\eta \circ \dX = \dX \circ \eta = \dX \circ \dX_\eta$, $\dX_\eta$ being a geodesic expansion in space-time mapping $\manif{W}_0$ into itself such that $\dX_\eta \circ X_0(\sigma) = X_0(f(\sigma))$, the integral measure must be the right invariant Haar measure of the group of geodesic expansions. Inserting \eqref{appl:metr} and \eqref{appl:ricci} into the generic form \eqref{geod:rmeas}, we find 
\begin{multline}
\label{appl:meas}
  \D_R \dX \simeq \prod_\sigma \left[ \sqrt{|h[X_0(\sigma)|]} \left(\frac{\sqrt{|g(\sigma)|}}{\N}\right)^\frac{D}{2} d^D \dX(\sigma)\right] \\
  \times \exp \left(-\frac{1}{6\N} \int d^d\sigma \sqrt{|g(\sigma)|}\; \tilde R_{\mu\nu}[X_0(\sigma)] \dX^\mu(\sigma) \dX^\nu(\sigma) \right),
\end{multline}
where $g$ is the determinant of the induced metric on $\manif{W}_0$. 
It is, of course, possible to derive the invariance of the integral measure \eqref{appl:meas} under $f_\eta$ directly from \eqref{appl:aut5}.

\subsection{Gauge Fixing}
\label{appl:gauge}
As established in the last subsection, the integral measure \eqref{appl:meas} is invariant under diffeomorphisms, which are represented by transformations $f_\eta$ (cf.\ Eqn.\ \eqref{appl:f}). Therefore, it contains the integration over unphysical degrees of freedom, which has to be separated. The standard method for this purpose is the Faddeev Popov procedure \cite{FadPop,Ramond}, which we shall apply here. 

Let us start by choosing a suitable gauge condition. First, decompose $\dX$ into components tangential and normal to $\manif{W}_0$,
\begin{equation}
\label{appl:xi}
  \xi_\alpha = \pa X_0^\mu h_{\mu\nu} \dX^\nu \qquad \text{and} \qquad 
  \xi_i = N_i^\mu h_{\mu\nu} \dX^\nu.
\end{equation}
For an immersed manifold of dimension $d$, the group of reparametrizations possesses $d$ degrees of freedom, which are given by the $d$ components of the vector field $\eta(\sigma)$. One could imagine to use these degrees of freedom to remove the $d$ tangential components $\xi_\alpha$ of $\dX$ by introducing the gauge condition $\xi_\alpha=0$. This is possible, if the gauge is accessible, i.e.\ if for any $\dX$ there exists a transformation $f_\eta$ such that $\pa X_0^\mu h_{\mu\nu} (f_\eta \circ \dX)^\nu=0$. This is indeed the case, if only in an approximative sense. For example, $\eta^\alpha = -\xi^\alpha$ yields the gauge condition to first order, but it is possible to construct a field $\eta$ containing terms up to an order, say, $n$, which yields the gauge condition to $n$-th order.  

The application of the Faddeev Popov procedure is straightforward, but there are some subtleties to be pointed out. First, let us introduce the inverse of the Faddeev Popov determinant by the usual expression \cite{Ramond} 
\begin{equation}
\label{appl:detdef}
  \Delta^{-1}[\dX] = \int \D f_\eta \; \delta_\mathrm{cov} \left[ \pa X_0^\mu h_{\mu\nu} \left(f_\eta \circ \dX \right)^\nu \right],
\end{equation}
where $\delta_\mathrm{cov}[\xi_\alpha]$ denotes the covariant delta functional satisfying $\delta_\mathrm{cov}[\xi_\alpha]=\delta_\mathrm{cov}[\xi^\alpha]$.
According to the standard procedure, the invariance property $\Delta[\dX]=\Delta[f_\eta\circ\dX]$ holds, if the measure $\D f_\eta$ is right-invariant, $\D f_\eta = \D (f_\eta \circ f_{\eta'})$. It is natural to identify $\D f_\eta = \D \eta$, but since $f_{\eta' \circ \eta} = f_\eta \circ f_{\eta'}$ due to Eqn.\ \eqref{appl:f}, the left-invariant Haar measure $\D_L \eta$ must be used. This is in perfect agreement with the result established in Sec.~\ref{geod:diff}, namely that the left-invariant Haar measure of the group of geodesic expansions is related to the integral over diffeomorphisms. As in Sec.~\ref{appl:riem} we can introduce a metric in the space of diffeomorphisms $f(\sigma)$. Then $\D_L \eta$ is obtained explicitely from \eqref{geod:measy2} as
\begin{multline}
\label{appl:etameas}
  \D f_\eta = \D_L \eta \simeq \prod_\sigma \left[ \sqrt{|g(\sigma)|} \left(\frac{\sqrt{|g(\sigma)|}}{\N}\right)^\frac{d}{2} d^d \eta(\sigma) \right] \\
  \times \exp \left[ \int d^d \sigma \frac{\sqrt{|g(\sigma)|}}{\N}
  \left(-\nabla_\alpha \eta^\alpha + \frac{1}{2} \nabla_\alpha \eta^\beta \nabla_\beta \eta^\alpha + \frac{1}{3} R_{\alpha\beta} \eta^\alpha \eta^\beta \right) \right]. 
\end{multline}

In order to calculate the Faddeev Popov determinant, we need an expression for the transformation of $\xi^\alpha$ under $f_\eta$. From \eqref{appl:aut5} and \eqref{appl:xi} we obtain 
\begin{align}
\notag
  \tord{\xi'}{}^\alpha &= \partial^\alpha X_0^\mu h_{\mu\nu} \tord{\dX'}{}^\nu \\
\label{appl:xiaut}
\begin{split}
  &=\xi^\alpha + \eta^\alpha + \eta^\beta \nabla_\beta \xi^\alpha + \frac{1}{2} \eta^\beta \eta^\gamma \nabla_\beta \nabla_\gamma \xi^\alpha - \frac{1}{2} \eta^\beta \eta^\gamma \xi^i \nabla_\beta H_{i\gamma}^\alpha - \frac{1}{6} \eta^\beta \eta^\gamma \eta^\delta H^{i\alpha}_\beta H_{i\gamma\delta} \\
  &\quad - H_{i\beta}^\alpha \left( \eta^\beta \xi^i + \eta^\beta \eta^\gamma \nabla_\gamma \xi^i + \frac{1}{2} \eta^\beta \eta^\gamma H^i_{\gamma\delta} \xi^\delta \right) + \frac{1}{3} \tilde R_{\mu\nu\lambda\rho} \partial^\alpha X_0^\mu \eta^\gamma \pc X_0^\rho \\
  &\quad \times \left[ N_i^\nu \xi^i \left(N_j^\lambda \xi^j + \pb X_0^\lambda \xi^\beta \right) + \pb X_0^\nu \left(\xi^\beta + \frac{1}{2} \eta^\beta \right) \left(N_i^\lambda \xi^i + \pd X_0^\lambda \xi^\delta \right) \right].
\end{split}
\end{align}
Solving \eqref{appl:xiaut} for $\eta$ to second order yields 
\begin{gather}
\label{appl:etasolv}
  \eta^\alpha \simeq \xi'{}^\alpha - \xi^\alpha - \left( \xi'{}^\beta -\xi^\beta \right) \nabla_\beta \xi^\alpha +  H_{i\beta}^\alpha \xi^i \left( \xi'{}^\beta - \xi^\beta \right),\\
\intertext{which reduces to}
\label{appl:etasolv2}
  \eta^\alpha \simeq - \xi^\alpha + \xi^\beta \nabla_\beta \xi^\alpha - H_{i\beta}^\alpha \xi^i \xi^\beta \qquad \text{for $\xi'{}^\alpha=0$.}
\end{gather}

Let us continue with the calculation of $\Delta[\dX]$. Using \eqref{appl:detdef} and \eqref{appl:etameas} and changing integration variables to $\xi'{}^\alpha$ we obtain
\begin{align}
\notag
\begin{split}
  \Delta^{-1}[\dX] &\simeq \int \prod_\sigma \Bigg[ \sqrt{|g(\sigma)|} \left(\frac{\sqrt{|g(\sigma)|}}{\N}\right)^\frac{d}{2} d^d \xi'(\sigma) \; \delta_\mathrm{cov}[\xi'{}^\alpha] \\
  & \quad \times \exp \left(-\nabla_\alpha \eta^\alpha + \frac{1}{2} \nabla_\alpha \eta^\beta \nabla_\beta \eta^\alpha + \frac{1}{3} R_{\alpha\beta} \eta^\alpha \eta^\beta \right) \det \left(\frac{\delta\xi'{}^\alpha}{\delta \eta^\beta}\right)^{-1} \Bigg] 
\end{split} \\
\label{appl:det1}
  &\simeq \prod_\sigma \left[ \exp \left(-\nabla_\alpha \eta^\alpha + \frac{1}{2} \nabla_\alpha \eta^\beta \nabla_\beta \eta^\alpha + \frac{1}{3} R_{\alpha\beta} \eta^\alpha \eta^\beta \right) \det \left(\frac{\delta \xi'{}^\alpha}{\delta \eta^\beta} \right)^{-1}\right]_{\xi'{}^\alpha=0}.\\
\intertext{Substituting \eqref{appl:etasolv2} for $\eta$, \eqref{appl:det1} yields}
\label{appl:det2}
\begin{split}
  \Delta[\dX] &\simeq \prod_\sigma \Bigg[ \exp \left( -\nabla_\alpha \xi^\alpha + \frac{1}{2} \nabla_\alpha \xi^\beta \nabla_\beta \xi^\alpha + \xi^\beta \nabla_\alpha \nabla_\beta \xi^\alpha \right. \\
  &\quad \left. - \nabla_\alpha \left( H^\alpha_{i\beta} \xi^\beta \xi^i \right) - \frac{1}{3} R_{\alpha\beta} \xi^\alpha \xi^\beta \right) \det \left(\frac{\delta \xi'{}^\alpha}{\delta \eta^\beta} \right)_{\xi'{}^\alpha=0} \Bigg].
\end{split}
\end{align}
The determinant appearing in \eqref{appl:det2} can be calculated to second order from \eqref{appl:xiaut}. Using also the geometrical relations from Sec.~\ref{appl:geom}, in particular Eqns.\ \eqref{geom:gauss2} and \eqref{geom:codazzi} we obtain
\begin{equation}
\label{appl:det3}
\begin{split}
  \det \left(\frac{\delta \xi'{}^\alpha}{\delta \eta^\beta} \right)_{\xi'{}^\alpha=0} &\simeq \exp \left( \nabla_\alpha \xi^\alpha - \frac{1}{2} \nabla_\alpha \xi^\beta \nabla_\beta \xi^\alpha - \xi^\beta \nabla_\alpha \nabla_\beta \xi^\alpha \right.\\
  &\quad + \nabla_\alpha \left( H^\alpha_{i\beta} \xi^\beta \xi^i \right) + \frac{1}{3} R_{\alpha\beta} \xi^\alpha \xi^\beta \\
  &\quad \left. -2 H_i \xi^i_0 - \frac{1}{2} H_{i\alpha\beta}H_j^{\alpha\beta} \xi^i \xi^j - \frac{1}{3} \tilde R_{\mu\nu\lambda\rho} \partial^\alpha X_0^\mu \pa X_0^\lambda N_i^\nu N_j^\rho \xi^i \xi^j \right),
\end{split}
\end{equation}
where 
\begin{equation}
\label{appl:xinought}
  \xi_0^i \simeq \xi^i - \xi^\alpha \nabla_\alpha \xi^i -\frac{1}{2} H^i_{\alpha\beta} \xi^\alpha \xi^\beta.
\end{equation}
Combining \eqref{appl:det2} and \eqref{appl:det3} finally yields 
\begin{equation}
\label{appl:det4}
  \Delta[\dX] \simeq \prod_\sigma \exp \left( -2 H_i \xi^i_0 - \frac{1}{2} H_{i\alpha\beta}H_j^{\alpha\beta} \xi^i \xi^j - \frac{1}{3} \tilde R_{\mu\nu\lambda\rho} \partial^\alpha X_0^\mu \pa X_0^\lambda N_i^\nu N_j^\rho \xi^i \xi^j \right).
\end{equation}

Before completing the Faddeev Popov procedure, let us look at the quantity $\xi^i_0$, which is defined by \eqref{appl:xinought}. From \eqref{appl:aut4} we obtain the transformation of the normal components $\xi^i$ under diffeomorphisms to second order,
\begin{equation}
\label{appl:xiiaut}
  \sord{\xi'}{}^i = \xi^i + \eta^\alpha \nabla_\alpha \xi^i +H^i_{\alpha\beta} \eta^\alpha \xi^\beta + \frac{1}{2} H^i_{\alpha\beta} \eta^\alpha \eta^\beta. 
\end{equation}
Using \eqref{appl:xiiaut} and the second order expression of \eqref{appl:xiaut}, we find that $\xi^i_0$ is (to second order) invariant under diffeomorphisms, $\xi'_0{}^i \simeq \xi_0^i$. Hence the Faddeev Popov determinant $\Delta[\dX]$ also is, as it should be, invariant under $f_\eta$. This is a good check and, in fact, would not be the case, if the integral measure $\D_R \eta$ were used instead of $\D_L \eta$. 

The last step of the Faddeev Popov procedure is strictly standard. Namely, for a functional $F$, which is invariant under diffeomorphisms, we obtain the integral
\begin{equation}
\label{appl:fadpop1}
  \int \frac{\D_R \dX}{\D f_\eta} F[\dX] = \int \D_R \dX \Delta[\dX] \delta_\mathrm{cov}[\xi^\alpha] F[\dX]. 
\end{equation}
It is easy to change integration variables from $\dX$ to its tangential and normal components $\xi$. Eqns.\ \eqref{appl:xi} yield
\begin{gather}
\notag
  \frac{\delta \dot X^\mu(\sigma)}{\delta \xi^\nu(\sigma')} = 
  \delta(\sigma,\sigma') A^{\;\mu}_\nu = \delta(\sigma,\sigma')
  \begin{cases}
    \pa X_0^\mu &\text{for $\alpha=\nu=1,\ldots,d$,}\\
	N_i^\mu     &\text{for $i=\nu=d+1,\ldots D$.}
  \end{cases}\\
\intertext{Thus we find}
\label{appl:dxxi} 
  A_\mu^{\;\rho} h_{\rho\lambda} A_\nu^{\;\lambda} = 
  \begin{pmatrix} g_{\alpha\beta} & 0\\ 0& \delta_{ij} \end{pmatrix}, 
  \qquad \det A = \sqrt{\frac{g}{h}}.
\end{gather}
Using \eqref{appl:dxxi} and the integral measure \eqref{appl:meas}, \eqref{appl:fadpop1} becomes after integration over the tangential components
\begin{equation}
\label{appl:fadpop2}
\begin{split}
  \int \frac{\D_R \dX}{\D f_\eta} F[\dX] &\simeq \int \D \xi \; F[N_i \xi^i] \exp \int d^d \sigma \frac{\sqrt{|g(\sigma)|}}{\N} \left( -2 H_i \xi^i - \frac{1}{2} H_{i\alpha\beta}H_j^{\alpha\beta} \xi^i \xi^j \right. \\
  &\quad \left. - \frac{1}{2} \tilde R_{\mu\nu\lambda\rho} \partial^\alpha X_0^\mu \pa X_0^\lambda N_i^\nu N_j^\rho \xi^i \xi^j - \frac{1}{6} \tilde R_{\mu\nu\lambda\rho} N^{k\mu} N_k^\lambda N_i^\nu N_j^\rho \xi^i \xi^j \right) 
\end{split}
\end{equation}
with the covariant integral measure
\begin{equation}
\label{appl:ximeas}
  \D \xi = \prod_\sigma \left[ \left( \frac{\sqrt{|g(\sigma)|}}{\N} \right)^{\frac{D-d}{2}} d^{D-d} \xi^i(\sigma) \right].
\end{equation}
Equation \eqref{appl:fadpop2} establishes our main result.

\section{Conclusions}
\label{concl}
After the semiclassical sum over immersed surfaces has been constructed we can ask, whether it can be used for a semiclassical treatment of, say the relativistic free particle or the bosonic string. The answer to this question is presumably positive, but more work needs to be done in order to achieve the set goal. Let us pinpoint the major problems. The first consists of the fact that the quantum path integral measure $[\D X]$ is in general not given by the ``naive'' measure $\D X$ of Eqn.\ \eqref{geod:meas}, but is obtained after integrating out the momenta of the phase space path integral. This is supported by the following argument. Consider the ``area'' (or Nambu-Goto) action 
\begin{equation}
\label{concl:S}
  S = \N^{-1} \int d^d \sigma \sqrt{|\det(\pa X^\mu \pb X^\nu h_{\mu\nu})|},
\end{equation}
whose semiclassical expansion is given by \cite{Vish96}
\begin{multline}
\label{concl:Sexp}
  S \simeq \N^{-1} \int d^d \sigma \sqrt{|g(\sigma)|} \left[ 1- 2 H_i \xi^i_0 \phantom{\frac12}\right. \\ \left. -\frac12 \xi_j \left( \delta_i^j \nabla^2 +  H^{j\alpha\beta} H_{i\alpha\beta} - 4 H^j H_i + \tilde R_{\mu\nu\lambda\rho} \partial^\alpha X_0^\mu \pa X_0^\lambda N^{j\nu} N_i^\rho \right) \xi^i \right].
\end{multline}
When calculating the partition function \eqref{intro:Z} we realize that, using the classical equation of motion $H^i=0$, all terms except the kinetic term and a term proportional to $\tilde R_{\mu\nu\lambda\rho}$ would cancel in the integral \eqref{appl:fadpop2}. Considering flat space-time, this cannot be the right result, since the kinetic term alone does not yield the critical dimension 26 for the bosonic string. Effective potential terms are needed, which will presumably stem from the phase space path integral. The term containing $\tilde R_{\mu\nu\lambda\rho}$ is less of a problem, because it is well known \cite{Kleinert} that path integrals in curved spaces need a quantum correction of exactly this form. It can thus be anticipated that all terms of this form cancel. 

Finally, we remark that the obtained results, which are in their present form correct to the lowest order of the semiclassical expansion, can in principle be extended to any desired order without changing the underlying qualities such as the group structure or the invariance properties of the integral measures. 

The author wishes to thank K.~S.~Viswanathan for many helpful conversations. This work was supported in part by an NSERC operating grant and a Graduate Fellowship at Simon Fraser University.

\renewcommand{\baselinestretch}{1} 

\end{document}